\documentclass[12pt,a4paper]{article}

\usepackage{epsf}
\begin{document}
\title{CAUSAL LOOP QUANTUM GRAVITY AND COSMOLOGICAL SOLUTIONS}
\author{Ali Shojai\thanks{shojai@theory.ipm.ac.ir} \& Fatimah Shojai\thanks{fatimah@theory.ipm.ac.ir}
\\ Department of Physics, University of Tehran, Tehran, Iran.}
\date{}
\maketitle
\begin{abstract}
We shall present here the causal interpretation of canonical quantum gravity in terms of new variables. 
Then we shall apply it to the minisuperspace of cosmology. A vacuum solution of quantum cosmology is obtained and the Bohmian trajectory is investigated. At the end a coherent state with matter is considered in the cosmological model.
\end{abstract}
\section{Introduction}
The causal interpretation, introduced by de-Broglie\cite{deb} and Bohm\cite{boh}, 
presents a definite trajectory for any quantum system.
It is proved that Copenhagen and causal interpretations answers to any \textit{common} question are  the same. There are questions 
\textit{specific} to causal interpretation, directly related to the trajectories which can not be asked in Copenhagen interpretation. Thus it is not simply another way of interpreting the same theory. But the problem that whether such questions are physical or not is still unclear\cite{zei}. 

The causal interpretation is motivated by the fact that the phase of the wave function for a non relativistic particle obeys the Hamilton--Jacobi equation modified by a potential called \textit{quantum potential}. 
Interpreting this equation as a quantum Hamilton--Jacobi equation one can derive the quantum trajectory. The power of this interpretation comes to the scene when one considers problems like measurement. It gives a deterministic version of measurement theory with the same results as the Copenhagen interpretation.  Extension of the theory to the case of relativistic particles\cite{rel} and  fields\cite{boh1} is straightforward.
It is also possible to make a causal interpretation of quantum gravity in terms of old variables\cite{old2}.

Here we shall investigate the possibility of making a causal interpretation of quantum gravity in terms of new variables\cite{ash}(loop quantum gravity).
\textit{Causal loop quantum gravity} may be useful at least in three points. The first point is the issue of time.
In loop quantum cosmology the notion of time can be introduced either internally, or externally.
In the first view the scalar constraint is looked at as an evolution equation with respect to the eigenvalues of momentum\cite{boj}. 
In the second view one introduces an external time which parameterizes quantum theory\cite{boj1}. It has no physical meaning but provides time dependent states, called \textit{state--time}\cite{boj1}.  
The time issue in Bohm's theory is different from these two definitions of time. In Bohmian mechanics the guidance relation connects canonical momentum to the gradient of the Hamilton--Jacobi function. Therefore one has a definite trajectory parameterized by this time, which is exactly the physical time for classical trajectory. In this case there is no approximation and if matter exists, one can eliminate the coordinate time, using the evolution of the matter fields, and introduce a clock field.
 
Another point is the evolution of the universe through the big bang singularity.
It is shown \cite{boj} that in loop quantum cosmology the quantum evolution is well defined through the big bang singularity. 
Investigation of existence or non--existence of singularity in causal loop quantum cosmology, means looking for any possible singularity in Bohmian trajectory. It is a general feature of de-Broglie--Bohm theory that it is quiet possible to have singularity in the wave function without any singularity in the trajectory.
  
The last point is about the role of the wave function.
Causal quantum theory can talk about single systems, like universe, without any reference to the probabilistic character of the wave function. Also in this theory, there is no need to have a classical system outside the observed system (the observer) to interpret the measurement problem. This shows the usability of the application of de-Broglie--Bohm theory to quantum cosmology.

In the next section we drive the guidance relation for loop quantum gravity and in section 3 we shall evaluate it for the cosmology minisuperspace.
In section 4 we shall find exact solutions and their Bohmian trajectories.
\section{Causal loop quantum gravity}
Loop quantum gravity is based on the formulation of general relativity in terms of $su(2)$ connections ($A_a^i$) and triads ($E^a_i$). Einstein's equations are expressed in terms of three constraints. Two of them (gauge and 3--diff constraints) are kinematical and the last one (the scalar or Hamiltonian constraint) is dynamical.

In loop quantum gravity the elementary variables are constructed from the above fields. They are holonomies along edges of a graph and smeared fluxes of triads through surfaces. These have quantum representations in the Hilbert space of the functions of holonomies.
An orthonormal basis for this kinematical Hilbert space is spin networks which are eigenvectors of geometric operators and depends on the connection via holonomies.

In order to have a causal interpretation of any quantum theory, one needs a guidance relation as well as the wave function.
The guidance relation enables one to derive Bohmian trajectories from the phase of the wave function. For the case of loop quantum gravity the wave functional is obtained by solving the above mentioned three constraints. The guidance relation can be obtained by using the Poisson bracket between holonomies and fluxes.
Therefore if we denote the holonomy along an edge $e$ as $A(e)$ and the electric flux through a surface $\Delta$ as $P(\Delta ,f)= 
\int_\Delta f_i\Sigma^i$ ($\Sigma^i_{ab}=\eta_{abc}P^{ci}$ is the 2-form dual of the electric field $P^a_i=E^a_i/8\pi G\gamma$, $\gamma$
is the Barbero--Immirizi parameter  and $f$ is a 
test function.), we have the commutation relation\cite{ash}:
\begin{equation}
\{ A(e),P(\Delta ,f) \} =-\frac{ 
\kappa(\Delta ,e)}{2}\left \{ 
\begin{array}{cc}
A(e)\tau^i f_i(p)&\textit{if p is the source of e}\\
-f_i(p)\tau^i A(e)&\textit{if p is the target of e}
\end{array}
\right .
\end{equation}
where $\kappa$ equals to +1(-1) if $e$ lies above (below) $\Delta$ and zero otherwise and $p$ is the point that the surface
$\Delta$ intersects with the edge $e$. 
Let us now derive the guidance equation. First consider a simple case when we have a Poisson bracket like $\{q,p\}=X(q)$. 
The quantum system would have the commutation relation $[\hat{q},\hat{p}]=i\hbar X(\hat{q})$, leading to the q-representation 
$\hat{p}=-i\hbar X\partial/\partial q$ and thus the Bohmian relation $p_{Bohm}=X\partial S/\partial q$. The extension to our case would be:
\begin{equation}
P(\Delta ,f)=-\frac{ 
\kappa(\Delta ,e)}{2} \left \{ 
\begin{array}{cc}
\textbf{trace}\left (\frac{\delta S}{\delta A(e)}A(e)\tau^i f_i(p)\right ) &\textit{if p is the source of e}\\
\textbf{trace}\left ( - \frac{\delta S}{\delta A(e)}f_i(p)\tau^i A(e)\right ) &\textit{if p is the target of e}
\end{array}
\right .
\label{guide}
\end{equation}
Thus in order to derive the Bohmian trajectories one should first solve the three quantum constraints and obtain the wave functional. Then denoting $\hbar$ times the phase of the wave functional as $S$, we can use the above guidance relation to obtain the quantum dynamics of the geometry. We shall do this for cosmological minisuperspace in the following section. 
\section{Causal loop quantum cosmology}
Using the spatial homogeneity and isotropy of cosmological models one can find solutions of the 3--diff and Gauss constraints\cite{boj}. The elementary variables are holonomies along straight edges and fluxes across squares. The kinematical Hilbert space is made from gauge invariant isotropic states which depends on the connection via the holonomies and form an orthonormal basis.

In such a minisuperspace the connection has only one degree of freedom $c=\frac{1}{2}(k-\gamma \dot{a})$ where $a$ is the scale factor and $k=0,+1$ is the curvature parameter for a flat and closed model respectively, and  the conjugate momentum is $p$ with $|p|=a^2$.
  
The action of scalar constraint after regularization and quantization on physical states leads to the following difference equation\cite{ash,boj}:
\[
(V_{\ell+5\ell_0}-V_{\ell+3\ell_0})e^{-2i\Gamma\ell_0}\psi(\phi,\ell+4\ell_0)
-\Omega(V_{\ell+\ell_0}-V_{\ell-\ell_0})\psi(\phi,\ell)+
\]
\begin{equation}
(V_{\ell-3\ell_0}-V_{\ell-5\ell_0})e^{2i\Gamma\ell_0}\psi(\phi,\ell-4\ell_0)=
-\frac{1}{3} 8\pi G\gamma^3\ell_0^3\ell_p^2\hat{C}_{matter}^{(\ell_0)}(\ell)\psi(\phi , \ell)
\label{diff}
\end{equation}
where $\Omega=2-4\ell_0^2\gamma^2\Gamma (\Gamma-1)$, $\ell$ is the eigenvalue of $p$, $\ell_0$ is a quantum ambiguity coming from 
regularization\cite{ash}, $\hat{C}_{matter}$ is the matter Hamiltonian, and  $\ell_p$  is the Planck length. $\Gamma$ is the spin connection related to the curvature parameter such that $\Gamma=0,1/2$ for the flat and close universe respectively.  
$\psi(\phi,\ell)$ are coefficients of the expansion
of the state in terms of spin network states:
\begin{equation}
\left | \psi \right\rangle =\sum_\ell \psi(\phi,\ell)\left| \ell \right\rangle
\end{equation}
in which  $\phi$ represents matter fields and $V_\ell=\left ( \frac{8\pi \gamma |\ell|}{6}\right )^{3/2}\ell_p^3$ is the eigenvalue of the volume operator. 
In order to the sum make sense, one should add the condition that $\psi(\phi,\ell)$ has support only on a countable subset of real numbers.
The gauge invariant isotropic state can be represented by 
$\langle c|\ell\rangle=e^{i\ell c}$ which form an orthonormal basis. The presence of $\ell_0$ is related to the quantum
ambiguity and it can be fixed by minimal area of the full theory.

As one may expect, the discrete evolution in terms of internal time $\ell$ does not break at big bang, this leads to resolution of singularity problem in quantum cosmology\cite{boj}. At scales enough larger than $\ell_0$, the differential WDW equation emerges as an approximation\cite{boj}. Also there is a lot of works done on effective field equation which incorporate quantum corrections to FRW equations\cite{boj3}.    

In order to have a causal interpretation, one needs the guidance relation. It can be derived from the guidance relation of the full theory given by the relation (\ref{guide}) by going to the minisuperspace of cosmology.  Taking $f_i=\tau_i$, since the holonomy for an edge of length $\ell$ is $A(e)=\cos\ell c/2+2\tau^i\sin\ell c/2$, using the relation (\ref{guide}), we have:
\begin{equation}
\frac{\delta S}{\delta c}=p
\end{equation}
In the next section we shall solve for the wave function in the absence of matter and derive the Bohmian trajectory.
\section{Cosmological solutions}
\subsection{Vacuum solution}
For simplicity here we shall restrict ourselves to vacuum solutions of equation (\ref{diff}). In the vacuum case the difference equation (\ref{diff}) can be written as:
\begin{equation}
F(\ell+4\ell_0)-\Omega F(\ell)+F(\ell-4\ell_0)=0
\end{equation}
where $F(\ell)=(V_{\ell+\ell_0}-V_{\ell-\ell_0})\psi (\ell)e^{-i\ell\Gamma/2}$.
Note that for $\ell=0$ this leads to $F=0$ and thus does not determine $\psi(0)$.
Using the Z--transform techniques, considering a solution of type $F(\ell)\sim e^{\beta \ell}$, the above equation leads to:
\begin{equation}
\beta=\left \{
\begin{array}{ccc}
\frac{ij\pi}{2\ell_0};&j\in Z;& k=0\\
\pm\beta_0;&\beta_0=\frac{\cosh^{-1}(1+\ell_0^2\gamma^2/2)}{4\ell_0};&k=1
\end{array}
\right .
\end{equation}
Since the Z--transform solution in the case $k=0$ is degenerate, one has a second solution of the form $\ell e^{\beta\ell}$. 
Therefore the general solution of equation (\ref{diff}) is:
\begin{equation}
\psi(\ell)=\left (\frac{6}{\gamma}\right )^{3/2}\frac{1}{\ell_p^3}
\frac{e^{i\Gamma\ell/2}}{|\ell+\ell_0|^{3/2}-|\ell-\ell_0|^{3/2}}
\left \{
\begin{array}{cc}
e^{ij\pi\ell/2\ell_0}(A+B\ell);&k=0\\
Ae^{\beta_0\ell}+Be^{-\beta_0\ell};&k=1
\end{array}
\right .
\end{equation}
where $A$ and $B$ are two constants. Note that this is only for $\ell\neq 0$, and that $\psi(0)$ is just another free parameter.

\begin{figure}[htb]
\begin{center}
\epsffile{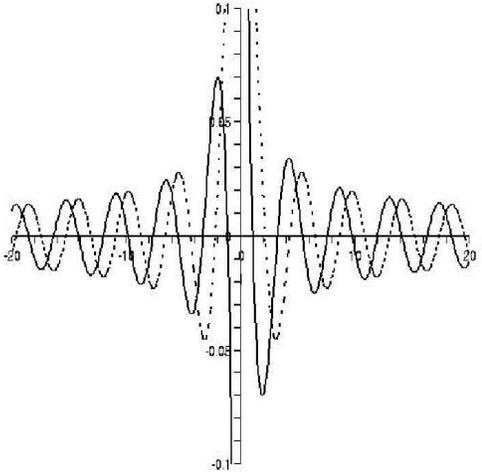}
\end{center}
\caption{\textit{\small The real (solid line) and imaginary (dashed line) parts of the wave function as a function of $\ell$. 
This is plotted for the case
$k=0$, $A=1$ and $B=0$.}}
\label{b1}
\end{figure}

\begin{figure}[htb]
\begin{center}
\epsffile{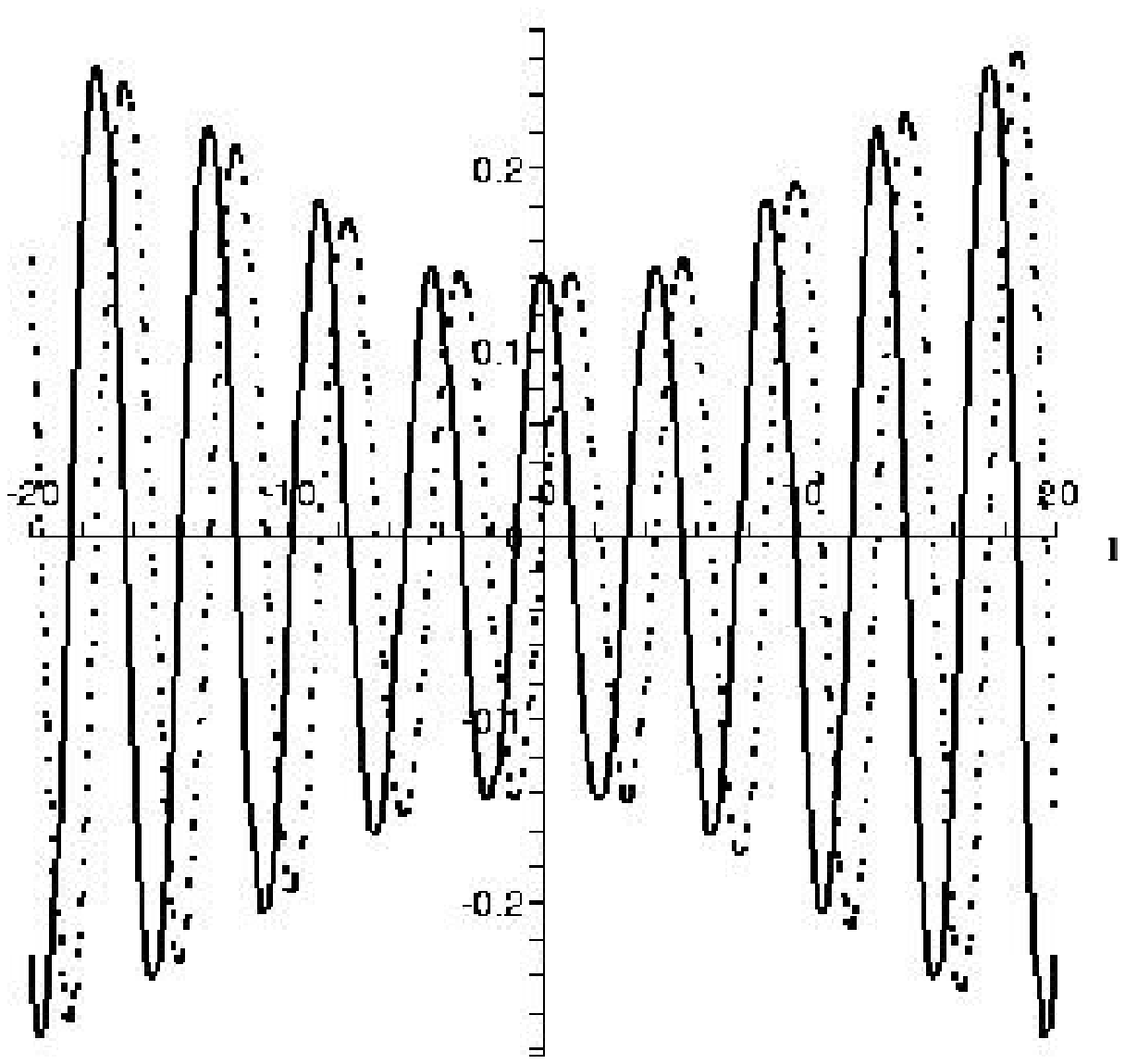}
\end{center}
\caption{\textit{\small The real (solid line) and imaginary (dashed line) parts of the wave function as a function of $\ell$. 
This is plotted for the case
$k=0$, $A=0$ and $B=1$.}}
\label{b2}
\end{figure}

\begin{figure}[htb]
\begin{center}
\epsffile{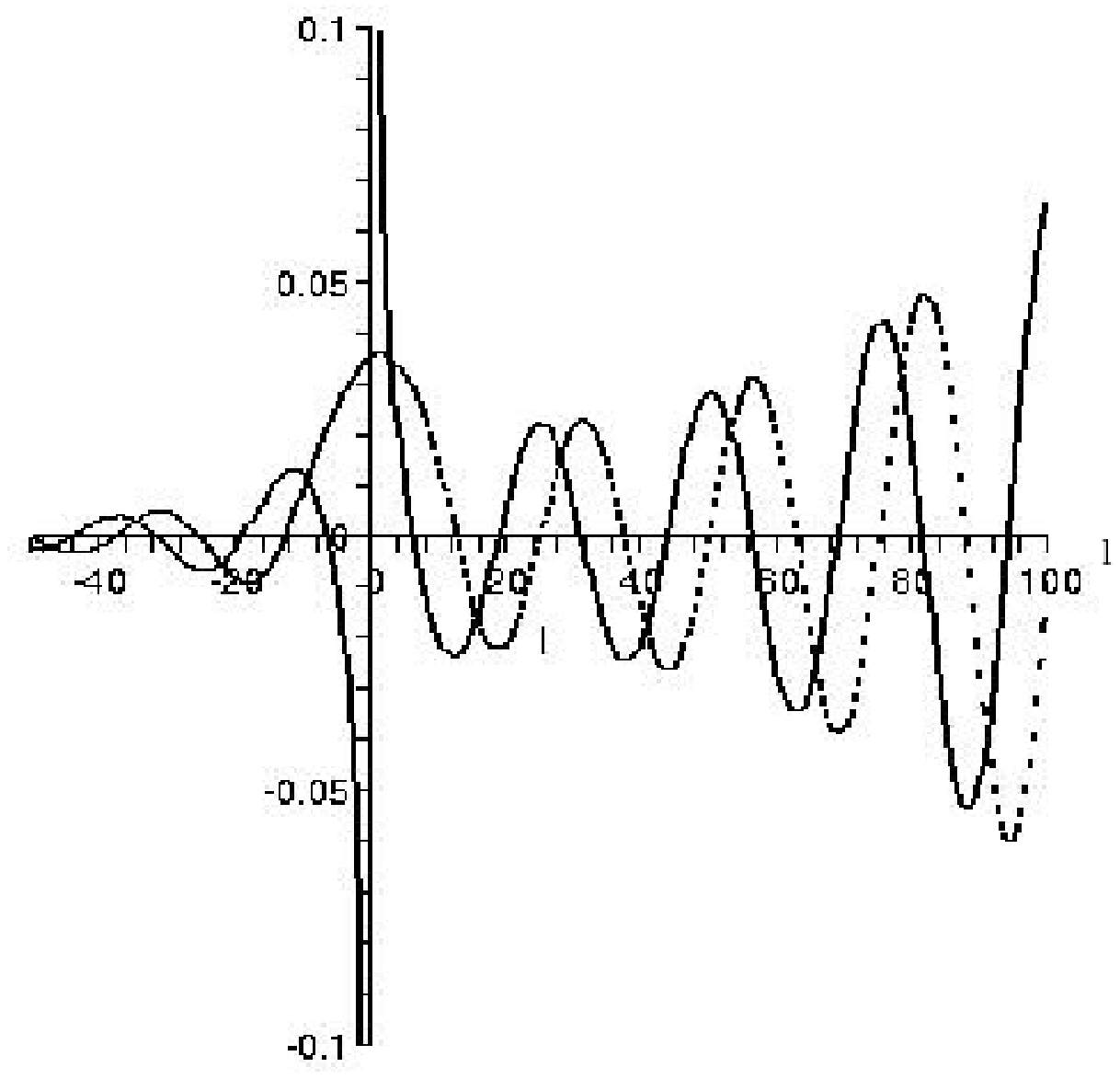}
\end{center}
\caption{\textit{\small The real (solid line) and imaginary (dashed line) parts of the wave function as a function of $\ell$. 
This is plotted for the case
$k=1$, $A=1$ and $B=0$.}}
\label{b3}
\end{figure}

\begin{figure}[htb]
\begin{center}
\epsffile{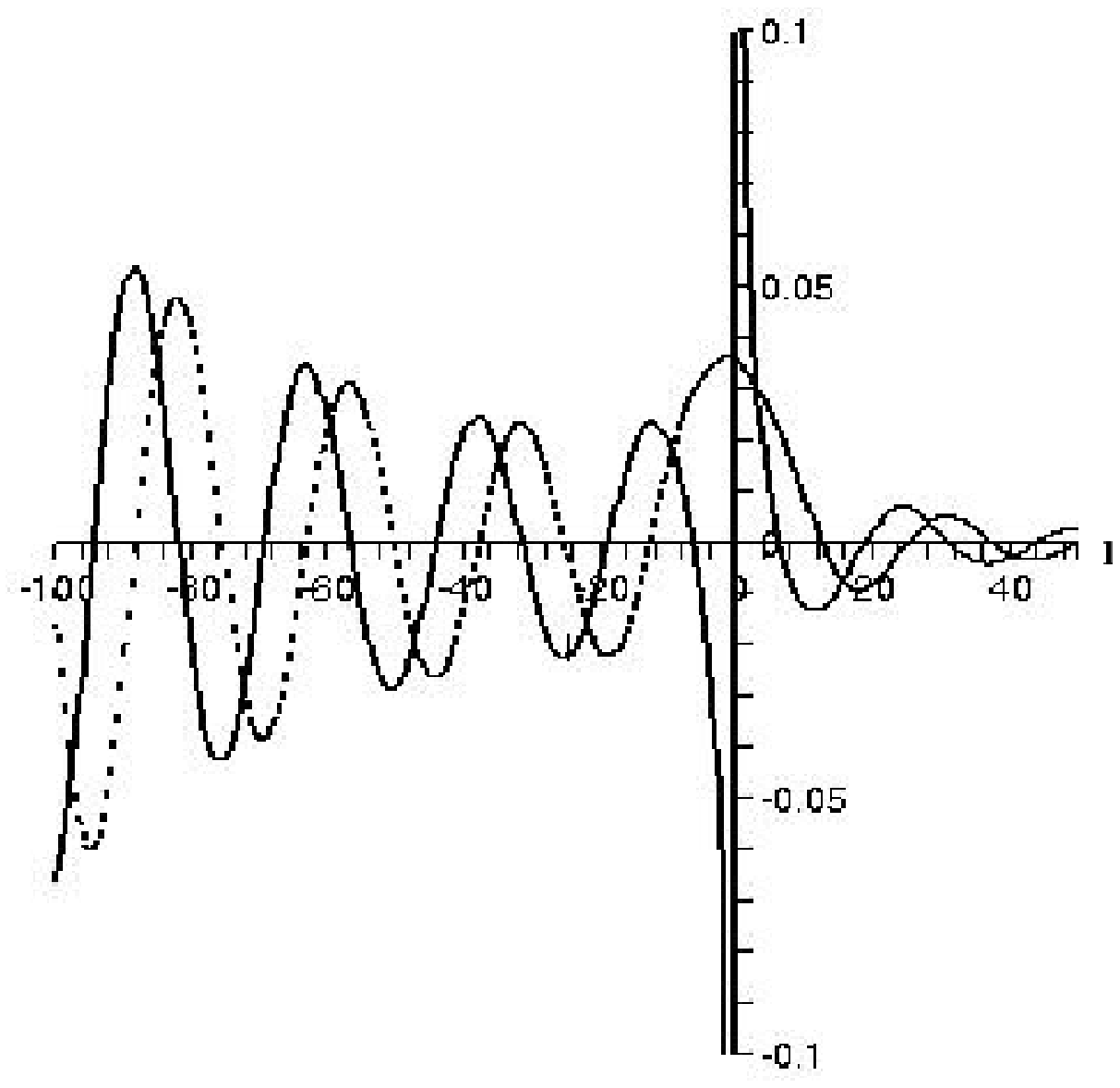}
\end{center}
\caption{\textit{\small The real (solid line) and imaginary (dashed line) parts of the wave function as a function of $\ell$. 
This is plotted for the case
$k=1$, $A=0$ and $B=1$.}}
\label{b4}
\end{figure}

For large $\ell$, we have the following asymptotic behavior:
\begin{equation}
\begin{array}{cc}
\psi(\ell)\sim \frac{1}{\sqrt{|\ell|}};&\textit{for } k=0,A\neq 0,B=0;\\
\psi(\ell)\sim \sqrt{|\ell|};&\textit{for } k=0,A=0,B\neq 0\\
\psi(\ell)\sim \frac{1}{\sqrt{|\ell|}}e^{\beta_0|\ell|};&\textit{for } k=1
\end{array}
\end{equation}
The behavior of the wave function is plotted in figures (\ref{b1})-- (\ref{b4}). It is clear from these figures that only for the case $k=0$ and $B=0$, the wave function converges for large $\ell$. The divergence in the other cases, is quiet similar to the results of \cite{unr}.
Here, thus we choose the convergent case.
As $\ell$ is a continuous variable, one can construct the general solution as the following form:
\begin{equation}
\left|\psi\right\rangle=\sum_{n_0\in [0,1)}
\sum_{n\in Z} f(n_0)\frac{e^{ij\pi(n+n_0)/2\ell_0}}{|n+n_0+\ell_0|^{3/2}-|n+n_0-\ell_0|^{3/2}}\left |\ell\right\rangle
\label{wf}
\end{equation}
where $f(n_0)$ is some coefficient, and again has support only on a countable subset. In the above relation we have broken the sum over the real number $\ell$ into sum over a unit interval and an integer $n$. The term $n_0=0$ in the above summation has the problem that setting $n=0$ leads to an infinite contribution. This suggests that one needs to introduce a condition on $f(n_0)$ as $f(0)=0$ and this makes a finite state.
\epsfxsize=3in \epsfysize=3in
\epsfxsize=3in \epsfysize=3in
\epsfxsize=3in \epsfysize=3in
\epsfxsize=3in \epsfysize=3in
\subsection{Bohmian trajectories}
In order to get some physical details from the wave function (\ref{wf}), we restrict ourselves to the case that $f(n_0)$ is equal to zero for all $n_0$ except a specific one, $\tilde{n}_0$. To derive the Bohmian trajectories, we use the wave function (\ref{wf}) in the configuration space. 
One can get an approximate closed form for the physical 
wave function noting that for large $|n|$ (i.e. grater than $|\tilde{n}_0+\ell_0|$):
\begin{equation}
|n+\tilde{n}_0+\ell_0|^{3/2}-|n+\tilde{n}_0-\ell_0|^{3/2}=3\ell_0 sgn(n)\sqrt{|n|}+{\cal O}(1/\sqrt{|n|})
\end{equation}
Introducing $\alpha=-\frac{j\pi}{2\ell_0}-c$, the wave function is approximately:
\begin{equation}
\left\langle c\left.\right|\psi\right\rangle \approx e^{-i\tilde{n}_0\alpha} \left \{ \frac{1}{(\tilde{n}_0+\ell_0)^{3/2}
-(\ell_0-\tilde{n}_0)^{3/2}} + \frac{1}{3\ell_0} \sum_{n=1}^\infty \frac{e^{-in\alpha}}{\sqrt{n}} -\frac{1}{3\ell_0}
\sum_{n=1}^\infty \frac{e^{in\alpha}}{\sqrt{n}}\right \}
\label{state}
\end{equation}
The last two terms can be summed to an integral form\cite{integral}:
\begin{equation}
-\frac{i\sin \alpha}{\Gamma(1/2)}\int_0^\infty 
\frac{dx}{\sqrt{x}(\cosh x -\cos\alpha)}=
-\frac{i}{2}\left [ e^{-i\alpha}\Phi(e^{-i\alpha},1/2,1)-e^{i\alpha}\Phi(e^{i\alpha},1/2,1)\right ]
\end{equation}
where $\Phi$ is the \textit{Lerch} $\zeta$--function $\Phi(z,s,v)=(\Gamma(s))^{-1}\int_0^\infty dt (t^{s-1}e^{-vt})/(1-ze^{-t})$.

\begin{figure}[htb]
\begin{center}
\epsffile{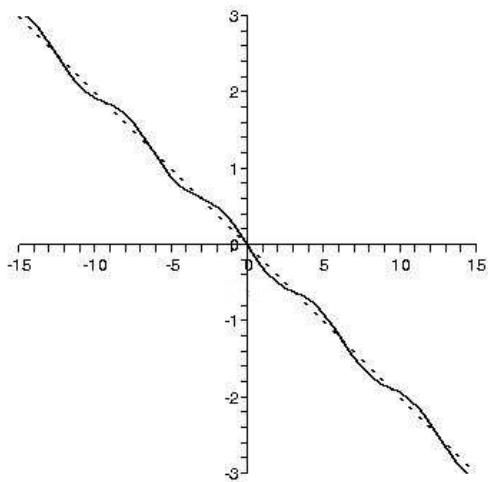}
\end{center}
\caption{\textit{\small Hamilton--Jacobi function (solid line is the quantum case and dashed line is the classical case) 
plotted versus $\alpha$.}}
\label{b5}
\end{figure}

\begin{figure}[htb]
\begin{center}
\epsffile{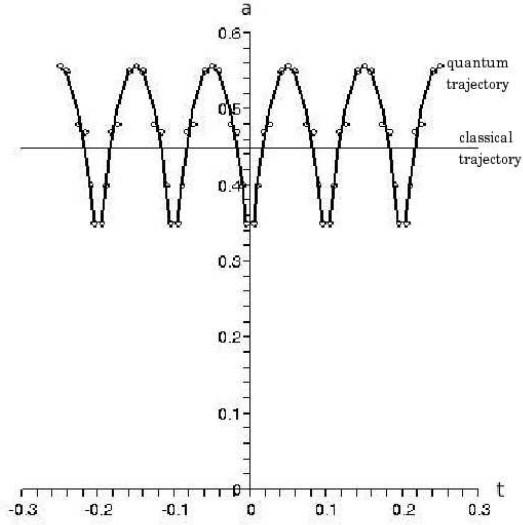}
\end{center}
\caption{\textit{\small Classical and quantum trajectories of the scale factor.}}
\label{b6}
\end{figure}

To drive the Bohmian trajectory one needs to extract the phase of the wave function in the configuration space.
This can be done numerically. The result is plotted in figure (\ref{b5}). In figure (\ref{b6}) the Bohmian trajectory for the scale factor is plotted. In these figures we have used $\ell_0=\sqrt{3}\pi$, $\gamma=0.238$ and $j=0$. As it is seen in figure (\ref{b5}), the quantum Hamilton--Jacobi function has some oscillations around the classical Hamilton--Jacobi function ($S_c=-\tilde{n}_0\alpha$).\footnote{This can be simply understood. Since the classical solution of such a model (FRW without matter) is a static universe ($a=$constant), we have $\delta S_c/\delta c=$constant and thus $S_c$ is a linear function in $c$.}
 As a result, Bohmian trajectory in figure (\ref{b6}) is also oscillatory around the classical path, which is a constant scale factor in this case.
\epsfxsize=3in \epsfysize=3in
\epsfxsize=3in \epsfysize=3in
\subsection{Inclusion of matter}
For simplicity we assume that the matter Hamiltonian is such that, an initial state which is highly peaked at the classical solution, would be remained so for a long time.
This condition is satisfied by choosing a Gaussian distribution like:
\begin{equation}
\left\langle c|\psi\right\rangle=\sum_n \psi(n,\phi)e^{inc}=\exp \left ( -\frac{(c-\tilde{c})^2}{2\sigma^2(\phi)}\right )
\end{equation}
Here $\tilde{c}$ is the classical connection and $\sigma$ is a complex function determining the width of the Gaussian, and we
denote its real and imaginary parts as:
\begin{equation}
\frac{1}{\sigma^2(\phi)}=\frac{1}{\sigma^2_1(\phi)}+\frac{i}{\sigma^2_2(\phi)}
\end{equation} 
The scalar constraint (\ref{diff}), determines the dependence of width $\sigma$ of the Gaussian distribution on matter.
Here we shall assume that the state is so peaked at the classical path that one can make a classical approximation for matter field (setting $\phi=\tilde{\phi}$ where $\tilde{\phi}$ is the classical matter field.). That is to say one can substitute matter fields in the guidance relation of the scale factor with its classical value.
The Hamilton-Jacobi function is then given by:\footnote{The fact that Hamilton--Jacobi function has this form for a Gaussian packet is a known character and can be found in the literature\cite{boh1}.}
\begin{equation}
S=-\frac{(c-\tilde{c})^2}{2\sigma_2^2(\tilde{\phi})}
\end{equation}
Using the guidance relation and the fact that $|p|=a^2$, the quantum path of the scale factor is given by the equation:
\begin{equation}
\dot{a}=\dot{\tilde{a}}\pm\frac{\sigma_2^2(\tilde{\phi})}{\gamma}a^2
\end{equation}
Therefore for such a semi--classical state the Bohmian trajectories are fluctuations around the classical path.

\textbf{Acknowledgment:} This work is supported by a grant from University of Tehran.

\end{document}